\documentstyle[12pt]{article}
\def\beq{\begin{equation}}
\def\eeq{\end{equation}}

\begin{document}

\begin{titlepage}
\begin{center}
{\Large \bf Theoretical Physics Institute \\
University of Minnesota \\}  \end{center}
\vspace{0.3in}
\begin{flushright}
TPI-MINN-92/45-T \\
September 1992
\end{flushright}
\vspace{0.4in}
\begin{center}
{\Large \bf Summing one-loop graphs at multi-particle threshold\\}
\vspace{0.2in}
{\bf M.B. Voloshin  \\ }
Theoretical Physics Institute, University of Minnesota \\
Minneapoilis, MN 55455 \\
and \\
Institute of Theoretical and Experimental Physics  \\
                         Moscow, 117259 \\
\vspace{0.2in}
{\bf   Abstract  \\ }
\end{center}

It is shown that the technique recently suggested by Lowell Brown for
summing the tree graphs at threshold can be extended to calculate the loop
effects. Explicit result is derived for the sum of one-loop graphs for the
amplitude of threshold production of $n$ on-mass-shell particles by one
virtual in the unbroken $\lambda \phi^4$ theory. It is also found
that the tree-level amplitude of production of $n$ particles by
two incoming on-mass-shell particles vanishes at the
threshold for $n > 4$.
\end{titlepage}

     The problem of calculating amplitudes of processes with many weakly
interacting particles has recently attracted a considerable interest,
initially triggered by the observation$^{\cite{ringwald,espinosa,mattis}}$
that such processes in particular are associated with a possible baryon and
lepton number violation in high-energy electroweak interactions.
Cornwall$^{\cite{cornwall}}$ and Goldberg$^{\cite{goldberg}}$ have pointed
out that in perturbative amplitudes with many external particles the weak
coupling may get compensated by large number of diagrams. This is a
manifestation of the old-standing problem of the factorial growth of the
coefficients in the perturbation theory$^{\cite{zinn}}$. Since the
perturbative expansion for multi-particle amplitudes starts from a high
order in the coupling constant, for sufficiently large number $n$ of
particles the factorial growth of the coefficients in the series
invalidates the perturbative calculation of such amplitudes. Given lack of a
better approach it seems useful to quantify and study the problem within the
perturbation theory itself. A simple model example in which the problem
arises with full strength is the amplitude $A_n$ where a virtual particle of
a real scalar field $\phi$ produces a large number $n$ of on-mass-shell
$\phi$-particles in the $\lambda \phi^4$ theory. It has been recently found
that the sum of all tree graphs for this amplitude in the threshold limit,
i.e. when all the produced particles are at rest, can be calculated
exactly for arbitrary $n$  both in the case of unbroken
symmetry$^{\cite{v}}$ and in the case of theory with spontaneous breaking of
the symmetry under the reflection $\phi \to -\phi~ ^{\cite{akp}}$.

Originally the calculation$^{\cite{v}}$ was done by directly solving a
recursion relation for the tree graphs. Argyres, Kleiss and
Papadopoulos$^{\cite{akp}}$ applied a regular method of solving the
recursion relations based on generating function for which the recursion
relation for the amplitudes $A_n$ is equivalent to second order non-linear
differential equation. Most recently Brown$^{\cite{brown}}$ has
shown that the generating function is nothing else than the classical field
$\phi_0(t)$ generated by an external source $\rho=\rho_0 e^{i mt}$ which
field is a complex solution of the Euler-Lagrange classical equation
satisfying the condition that it has only the positive frequency part. The
equations and their solutions in both approaches are related by a simple
change of variable. Thus Brown has reproduced the previous
results$^{\cite{v,akp}}$ in a simple and elegant way.

The purpose of this paper it to show that Brown's technique can be extended
to calculate the loop contributions to the amplitudes $A_n$ as well as the
amplitudes of more complicated processes e.g. of the scattering $2 \to n$.
For definiteness the case of unbroken $\lambda \phi^4$ theory with the
Lagrangian

\beq
{\cal L} = {1 \over 2} (\partial \phi)^2 -{1 \over 2} m^2 \phi^2 - {1 \over
4} \lambda \phi^4
\label{lagrangian}
\eeq
will be considered and we will concentrate on the calculation of the
one-loop correction to the threshold amplitudes $A_n$. The result of this
calculation can be written as

\beq
\langle 2k+1|\phi(0)|0\rangle  = (2k+1)! \left ( {{\bar
\lambda} \over {8 {\bar m}^2}} \right )^k \left [ 1-k(k-1){{ 3^{3/2}
\lambda} \over {16 \pi^2}} \left ( \ln {{2+\sqrt{3}} \over {2- \sqrt{3}}} -
i~ \pi \right ) \right ]~~,
\label{result}
\eeq
where ${\bar \lambda}$ and ${\bar m}$ are the renormalized coupling
constant and the mass, the appropriate renormalization condition, which
specifies the finite terms, will be described below. Equation (\ref{result})
gives the exact sum of all tree and one-loop graphs at the threshold of
production of $n=2k+1$ $\phi$-particles for arbitrary  $k$ {\footnote {The
number $n$ of final particles produced by one virtual is necessarily odd
due to the unbroken reflection symmetry.}}.
The formula (\ref{result}) gives the relative magnitude of the
first loop correction growing as $n^2$ at large $n$. This behavior
explicitly demonstrates invalidity of the previous
arguments$^{\cite{v2,v3}}$ of the present author that terms, containing $n^2
\lambda$ should be absent in the loop effects. The reason for those
arguments to be faulty is related to the singularities of the underlying
classical field in the plane of complex time, where the quantum fluctuations
are more singular than the classical background. However, the presence of
the $n^2 \lambda$ parameter in the loop effects does not necessarily imply
that the quantum effects completely eliminate the growth of the amplitudes
and a further study is needed. The relatively easy calculability of the
threshold amplitudes at the tree and the one-loop level and the remarkably
simple form of the result (\ref{result}) gives us a hope that these
amplitudes can be studied well beyond the first terms in the perturbative
expansion.

That the amplitudes in the $\lambda \phi^4$ theory at the multi-particle
thresholds may have special properties is also hinted at by another fact,
which follows as a by-product from the calculation in this paper. Namely, if
one considers the sum of all tree graphs for the amplitude of the process
where two incoming on-shell particles produce $n$ on-shell particles exactly
at the threshold, it turns out that this sum is non-zero only for $n=2$ and
$n=4$ and is vanishing for all $n > 4$ (the number of final particles in
this case is necessarily even). It is in fact due to this behavior that the
one-loop term in eq.(\ref{result}) contains only the 4-particle threshold
factor.

The technique suggested by Brown$^{\cite{brown}}$ is based on the standard
reduction formula representation of the amplitude through the response of
the system to an external source $\rho(x)$, which enters the term $\rho
\phi$ added to the Lagrangian

\beq
\langle n|\phi(x)|0\rangle  =\left [ \prod_{a=1}^n \lim_{p_a^2 \to m^2} \int
d^4 x_a~ e^{i p_a x_a} (m^2-p_a^2)~ {\delta \over {\delta \rho(x_a)}} \right
] \langle 0_{out}|\phi(x)|0_{in}\rangle ^\rho |_{\rho=0}~~,
\label{reduc}
\eeq
the tree-level amplitude being generated by
the response in the classical approximation, i.e. by the classical solution
$\phi_0(x)$ of the field equations in the presence of the source.

For all the spatial momenta of the final particles equal to zero it is
sufficient to consider the response to a spatially uniform time-dependent
source $\rho(t) = \rho_0(\omega) e^{i\omega t}$ and take the on-mass-shell
limit in eq.(\ref{reduc}) by tending $\omega$ to $m$. The spatial integrals
in eq.(\ref{reduc}) then give the usual factors with the normalization
spatial volume, which as usual is set to one, while the time dependence on
one common frequency $\omega$ implies that the propagator factors and the
functional derivatives enter in the combination

\beq
(m^2-p_a^2){\delta \over {\delta \rho(x_a)}} \to (m^2-\omega^2) {\delta
\over {\delta \rho(t)}}={\delta \over {\delta z(t)}} \label{zt}~~, \eeq
where

\beq
z(t)={{\rho_0(\omega) e^{i \omega t}} \over {m^2-i \epsilon -\omega^2}}
\label{zl}~~
\eeq
coincides with the response of the field to the external source in the limit
of absence of the interaction, i.e. of $\lambda =0$. For a finite amplitude
$\rho_0$ of the source the response $z(t)$ is singular in the limit $\omega
\to m$. The crucial observation of Brown$^{\cite{brown}}$ is that, since
according to eq.(\ref{zt}) we need the dependence of the response of the
interacting field $\phi$ only in terms of $z(t)$, one can take the limit
$\rho_0(\omega) \to 0$ simultaneously with $\omega \to m$ in such a way that
$z(t)$ is finite,

\beq
z(t) \to z_0 e^{imt}~~.
\label{z0}
\eeq

Furthermore, to find the classical solution $\phi_0(x)$ in this limit one
does not have to go through this limiting procedure, but rather consider
directly the on-shell limit with vanishing source. The field equation with
zero source is of course given by

\beq
\partial^2 \phi + m^2 \phi + \lambda \phi^3 =0~~.
\label{el}
\eeq
For the purpose of calculating the matrix element in eq.(\ref{reduc}) at
the threshold one looks for a solution of this equation which depends only
on time and contains only the positive frequency part with all harmonics
being multiples of $e^{imt}$. The solution satisfying these conditions reads
as$^{\cite{brown}}$:

\beq
\phi_0(t)={{z(t)} \over {1-(\lambda/8 m^2) z(t)^2}}~~
\label{phi0}
\eeq

According to equations (\ref{zt}) and (\ref{reduc}) the $n$-th derivative of
this solution with respect to $z$ gives the matrix element $\langle
n|\phi(0)|0\rangle $ at the threshold in the tree approximation:

\beq
\langle 2k+1|\phi(0)|0\rangle _0= \left ( {\partial \over {\partial z}}
\right )^{2k+1} \phi_0 |_{z=0}=(2k+1)!  \left ( {\lambda \over {8
m^2}} \right )^k~~,
\label{a0}
\eeq
which reproduces the previously known result$^{\cite{v}}$.
The fact that the matrix element is non-zero only for odd $n$ obviously
follows from that the expansion of $\phi_0$ in eq.(\ref{phi0}) contains
only odd powers of $z$.

It can be noticed that the solution (\ref{phi0}) is in fact not uniquely
determined by the above mentioned conditions. Namely, $z(t)$ can be rescaled
by an arbitrary constant $C$. This constant corresponds to the choice of
normalization of the field, so that the value $C=1$ is fixed by the usual
normalization condition $\langle 1|\phi(0)|0\rangle =1$, as can be seen from
the linear term in the expansion of $\phi_0$ in powers of $z$.

Another important point concerning the solution (\ref{phi0}) is related to
the fact that this solution is essentially complex for real time $t$. This
is imposed by the that in calculating production of particles by the virtual
field, rather than both production and absorption, one necessarily has to
consider only the positive frequency part of the field, which is essentially
complex.

The quantum loop corrections to the amplitudes $\langle n|\phi|0\rangle $
are obtained by substituting instead of the classical field the mean value
of the full field

\beq
\phi(x)=\phi_0(x)+\phi_q(x)~~,
\label{quant}
\eeq
where $\phi_q(x)$ is the quantum part of the field. Expanding the field
equation (\ref{el}) near the classical solution $\phi_0$ and retaining only
the first non-vanishing quantum correction, one finds that the mean field
$\phi(x)$ to the first quantum order satisfies the equation

\beq
\partial^2 \phi(x) + m^2 \phi(x) + \lambda \phi(x)^3 + 3 \lambda \phi_0(x)
\langle \phi_q (x) \phi_q (x)\rangle =0~~,
\label{qel}
\eeq
where $\langle \phi_q (x) \phi_q (x)\rangle $ is the limit of the
Green function in the classical background field $\phi_0$

\beq
G(x_1,x_2)=\langle T(\phi_q (x_1) \phi_q (x_2))\rangle
\label{gf}
\eeq
when its arguments are at the same point $x$.

Therefore the steps needed to calculate the first loop correction to the
amplitudes $A_n$ are the following:  \\
{\it i}. Calculate the Green function (\ref{gf}) as the inverse of the
operator of the second variation of the action,

\beq
\partial^2 +m^2 +3\lambda \phi_0(x)^2~~;
\label{oper}
\eeq
{\it ii}. Find its limit in coinciding points, which enters
eq.(\ref{qel});\\
{\it iii}. Expand the solution of thus found equation in powers of $z(t)$,
which gives the amplitudes at the threshold in the same way as in
eq.(\ref{a0}).

This program however is obscured at the very first step by the fact that
with essentially complex $\phi_0$ (eq.(\ref{phi0})) the operator
(\ref{oper}) is essentially non-Hermitean. However one can render this
operator real
and thus the problem more tractable by analytical
continuation in time $t$, which amounts to rotation and shift in the complex
plane. Namely, the substitution which achieves the goal reads as

\beq
\sqrt{{\lambda \over {8 m^2}}} z(t) = i e^{m \tau}
\label{trans}
\eeq
and the variable

\beq
\tau=i t + {1 \over m} \ln {{\lambda z_0} \over {8 m^2}} -{{i \pi} \over
{2m}}
\label{tau}
\eeq
is then used as the new time variable $t$. The
necessity of the shift in addition to the usual rotation to the Euclidean
time is caused by existence of a pole of $\phi_0(t)$ on the negative
imaginary axis, where the operator (\ref{oper}) is singular. The poles are
repeated parallel to the real axis with the period $\pi/m$.
The axis, corresponding to real $\tau$, on which the operator (\ref{oper})
is real
runs parallel to the imaginary axis of $t$ exactly in the middle
between two poles, see Fig.1. It should be emphasized however that it is the
pole structure of the field, which gives rise to the factorial growth of the
multi-particle amplitudes the study of which may eventually be the central
point in solving the problem of multi-boson processes. Here, for the
purpose of the specific calculation, we chose to stay away from the poles to
avoid explicit singularity in the equations.

To somewhat simplify the notation we set the mass $m$ equal to one and
restore it when needed and also introduce the notation
$u(\tau)=e^\tau=-i \sqrt{\lambda /8 }z(t)$. For real $u(\tau)$ the classical
field (\ref{phi0}) is purely imaginary:

\beq
\phi_0(u(\tau))= \sqrt{8 \over \lambda} {{i u} \over {1+u^2}} = \sqrt{2
\over \lambda} {i \over {{\rm cosh}~\tau}}
\label{xt}
\eeq
and the operator (\ref{oper}) is real. In a mode with spatial momentum ${\bf
k}$ the operator has the form

\beq
-{ {d^2} \over {d \tau^2}} + \omega^2 -{6 \over {( {\rm cosh}~\tau)^2}}~,
\label{emode}
\eeq
which is the familiar operator in one of exactly solvable potentials in
Quantum Mechanics (see e.g. Ref.\cite{ll}), and $\omega$ is the energy of
the mode: $\omega^2= {\bf k}^2+1$.

The regular at $\tau \to +\infty$ solution of the homogeneous equation with
the operator (\ref{emode}) has the form

\beq
f_1(u(\tau))={{2 - 3\,\omega + {\omega^2} - 8\,{u^2} + 2\,{\omega^2}\,{u^2}
+ 2\,{u^4} + 3\,\omega\,{u^4} +
{\omega^2}\,{u^4}}\over {{u^\omega}\,{{\left( 1 + {u^2} \right) }^2}}}
\label{f1}
\eeq
and the solution regular at $\tau \to -\infty$ is given by

\beq
f_2(u(\tau))=f_1(1/u(\tau))={{{u^\omega}\,\left( 2 + 3\,\omega + {\omega^2}
- 8\,{u^2} + 2\,{\omega^2}\,{u^2} + 2\,{u^4} -
 3\,\omega\,{u^4} + {\omega^2}\,{u^4} \right) }\over
{{{\left( 1 + {u^2} \right) }^2}}}~~.
\label{f2}
\eeq
The Wronskian of these solutions is given by

\beq
W=f_1(\tau) f_2^\prime (\tau)-
f_1^\prime (\tau) f_2(\tau)= 2 \omega (\omega^2-1) (\omega^2-4)~~.
\label{wr}
\eeq
The convention for the sign of the Green function used here is specified
by the explicit expression for the Green function in partial wave
with the spatial momentum ${\bf k}$ in terms of $f_1,~f_2$ and $W$:
$G_\omega (\tau_1,\tau_2)=f_1(\tau_1) f_2(\tau_2)/W$ for $\tau_1 >  \tau_2$
and $G_\omega (\tau_1,\tau_2)=f_1(\tau_2) f_2(\tau_1)/W$ for $\tau_2 >
\tau_1$.

Naturally, having the explicit expression for the Green function one can
also evaluate amplitudes of more complicated processes, say,
the tree level amplitude of
the threshold production of $n$ particles by two incoming on-mass-shell
particles of high energy. However equations (\ref{f1}), (\ref{f2}) and
(\ref{wr}) show that there is in fact almost nothing to calculate for the
latter amplitude:  the Green function has poles only at $\omega^2=1$ and
$\omega^2=4$ (the zeros of the Wronskian (\ref{wr})). By the reduction
formula this implies that the on-mass-shell amplitude is non-vanishing only
at these values of the energy of each of the two incoming particles.  The
case $\omega=1$ corresponds to the trivial process $2 \to 2$ at the
threshold, while the case $\omega=2$ corresponds to the process $2 \to 4$.
(In the rest frame of the produced particles, which is used throughout this
paper, $\omega$ corresponds to the energy of each of the two incoming
particles, so that the total energy is $2\omega=4$.) The absence of other
poles of the Green function at higher $\omega$ means that for $n > 4$ the
sum of tree graphs for the on-shell process $2 \to n$ is vanishing.

After this remark we proceed with calculating the loop
correction in eq.(\ref{result}).
The partial wave Green function at coinciding points is given by

\beq
g_\omega (\tau) = G_\omega (\tau, \tau)=f_1(\tau) f_2(\tau)/W
\label{gtt}
\eeq
which yields the average value of the square of quantum fluctuations after
integrating over ${\bf k}$:

\beq
\langle \phi_q(\tau) \phi_q(\tau)\rangle  = g(\tau) = \int g_\omega (\tau)
{{d^3 k} \over {(2 \pi)^3}} = {1 \over {2 \pi^2}} \int_1 g_\omega (\tau)
\omega \sqrt{\omega^2-1} ~d\omega ~.
\label{intom}
\eeq
The calculation of the latter integral involves problems related to the
on-shell singularities and to the ultra violet divergence. The on-shell
singularities correspond to the zeros of the Wronskian (\ref{wr}) at
$\omega^2=1$ and $\omega^2=4$. The first of these corresponds to the
translational zero mode of the classical solution $\phi_0$ and in fact
produces no effect in the integral in eq.(\ref{intom}) since the singularity
at $\omega^2=1$ is integrable. (This is why in the 4-dimensional theory one
does not have to consider subtraction of the contribution of the zero mode
from the Green function.) The pole at $\omega^2=4 m^2$ (the dependence
on mass is restored) is dealt with using the Feynman's $i \epsilon$ rule
i.e. by shifting the pole to the negative half-plane $m^2 \to
m^2-i\epsilon$. The integral then develops imaginary part, which in the end
corresponds to the dynamical imaginary part of the one-loop graphs, dictated
by the unitarity.

To separate the ultra violet divergent terms we expand $g_\omega (\tau)$ in
powers of $\omega^{-1}$ and find that the two terms, which give the
quadratic and the logarithmic divergence have the form

\beq
g_\omega (\tau) = {1 \over {2 \omega}}+ {{6 u^2} \over {(1+u^2)^2
\omega^3}}+g_\omega^r (\tau)~,
\label{expan}
\eeq
where the regular part $g_\omega^r (\tau)$ contains terms of the order
$\omega^{-5}$ and higher, so that its contribution to the integral in
eq.(\ref{intom}) is finite in the ultra violet. After this decomposition the
result of the integration in eq.(\ref{intom}) can be presented as

\beq
g(\tau)={1 \over 2} I_1 + {{6 u^2} \over {(1+u^2)^2 }} (I_3+{1 \over {2
\pi^2}}) - {{6 u^4} \over {(1+u^2)^4}} F~,
\label{gff}
\eeq
where

\beq
F={\sqrt{3}
 \over {2 \pi^2}} \left ( \ln {{2+\sqrt{3}} \over {2- \sqrt{3}}} -
i~ \pi \right )
\label{fac}
\eeq
and $I_1,~I_3$ are the ultra violet divergent integrals:

\beq
I_n= {1 \over {2 \pi^2}} \int  \omega^{1-n} \sqrt{\omega^2 -1} ~d\omega~.
\label{divint}
\eeq
In equation (\ref{gff}) we have also combined with the logarithmically
divergent integral a part of the finite contribution from integration of the
$g_\omega^r (\tau)$, which has the same functional dependence on $u(\tau)$,
hence the factor $(I_3 +1/(2 \pi^2))$.

The divergent contributions can be regularized in a standard way, the most
straightforward being the Pauli-Villars regularization. Upon substitution
into equation (\ref{qel}) for the mean field with the quantum correction
the quadratically divergent part proportional to $I_1$ gives rise to a term
linear in the classical field $\phi_0$ while the logarithmically divergent
part proportional to $I_3$ results in a correction to the term with
$\lambda \phi_0^3$. Therefore these terms can be dumped into the definition
of the renormalized mass ${\bar m}$ and the coupling constant ${\bar
\lambda}$ according to

\begin{eqnarray}
{\bar m}^2=m^2+{{3 \lambda } \over 2} I_1 \nonumber \\
{\bar \lambda}=\lambda - {{9 \lambda^2} \over 4} (I_3 + {1 \over {2
\pi^2}})~.
\label{renorm}
\end{eqnarray}
These definitions can be used to
relate the quantities ${\bar m}$ and ${\bar \lambda}$ to the renormalized
constants in any other renormalization scheme.  One can readily see that the
divergent parts are scheme-independent, while the relation between the
finite parts depends on the specific definition of the regularization
procedure.

The only non-trivial modification of the average field given by equation
(\ref{qel}) is related to the finite part of the average value of the square
of quantum fluctuations (eq.(\ref{intom})), proportional to the constant
factor $F$. If one seeks the solution of the equation (\ref{qel}) in the
form $\phi(t)=\phi_0(t;{\bar m},{\bar \lambda})+\phi_1(t)$, where the
renormalization of the constants is plugged into the functional dependence
of the classical solution, the equation for the correction $\phi_1(\tau)$
(i.e. on the $\tau$ axis) reads as

\beq
\left ( {{d^2} \over {d \tau^2}}  - 1 + {{24 u^2} \over {(1+u^2)^2}} \right
) \phi_1 = -i 18 \lambda \sqrt{8 \over \lambda} F {{u^5} \over
{(1+u^2)^5}}~,
\label{eq1}
\eeq
the condition on the appropriate solution to this equation being that its
expansion in $u$ starts with the fifth power, since only starting from final
states with five particles the threshold amplitudes develop an imaginary
part, which in this calculation originates in the imaginary part of $F$.
The solution satisfying this condition is

\beq
\phi_1(\tau)=-i {{3 \lambda} \over 4} \sqrt{8 \over \lambda} F {{u^5} \over
{(1+u^2)^3}}~,
\label{phi1}
\eeq.
Using equation (\ref{trans}) one can readily restore from here the response
of the field in terms of $z(t)$ with the first quantum correction included:

\beq
\phi_{0+1}(t)= {{z(t)} \over {1-({\bar \lambda}/8 {\bar m}^2) z(t)^2}} \left
(1-{{3 \lambda} \over 4} F {{(\lambda /8 m^2)^2 z(t)^4} \over {(1-(\lambda
/8 m^2) z(t)^2)^2 }} \right )
\label{phi01}
\eeq
and by expanding in series in $z(t)$ finally arrive at the result in
equation (\ref{result}).

The rotation (\ref{trans}) used here may invite the objection, that such
rotation in the path integral is obstructed by the infinite chain of poles
parallel to the real axis of $t$, which may give rise to extra contributions
in the quantum effects. However it can be explicitly shown that this does
not happen at least at the one-loop level. Namely, it is a straightforward
(but rather cumbersome) exercise to verify that the recursion relations for
the sum of graphs for the propagator of the field $\phi$ with emission of
$n$ on-shell particles all being at rest are equivalent to the differential
equation for the Green function of the operator (\ref{emode}) and then that
the recursion relations for the loop graphs are equivalent to the equation
(\ref{eq1}) on the $\tau$ axis. Another simple (and in no way rigorous)
check is to verify the formula (\ref{result}) for few first $n$ by direct
computation of the graphs. This also turned to be helpful in checking the
relative coefficients and signs in the equations of this paper.
The remarkably simple form of the result (\ref{result}) suggests that there
may be a way to calculate further quantum effects. In particular one can
notice that the finite term, proportional to the factor $F$ (eq.(\ref{fac}))
has the form given by the simple scalar vacuum polarization at $q^2=16m^2$.
This of course is a consequence of the eigenmode of the operator
(\ref{emode}) at $\omega^2=4$, or, equivalently, of the fact that the
tree-level threshold amplitudes of the processes $2 \to n$ are equal to
zero for $n > 4$.  In terms of the graphs the cancellation of the
contributions to the imaginary and the real parts of the thresholds at
higher $q^2$ looks quite surprising.

In the present calculation we have avoided approaching the poles of the
classical solution $\phi_0$, where the quantum expansion in fact breaks
down, since the quantum fluctuations are more singular than the classical
solution. However those are the singularities of the field in the complex
plane of $t$ (or equivalently of $z$) which give rise to the factorial
growth of the amplitudes. The appearance of the singularity at the imaginary
axis follows from the simple fact that on this axis the classical field
equation

\beq
{{d^2} \over {dt^2}} \phi = m^2 \phi + \lambda \phi^3
\label{imt}
\eeq
corresponds to the free fall in the inverted $\lambda \phi^4$ potential
which takes a finite time for a finite staring value $\phi(0)$. It looks
at least extremely unnatural that quantum effects would slow down this fall
to the extent that the time of the fall would be infinite. For any finite
time, however, the singularity of the field there will produce the factorial
growth of the amplitudes.

As a simple final remark, it can be mentioned that though the present
calculation is done for the case of unbroken symmetry it looks quite
straightforward to apply the same technique to the case of the spontaneously
broken symmetry.

I am thankful to Lowell Brown for sending to me the preprint of his paper
and to Arkady Vainshtein for stimulating discussions. I also acknowledge an
extensive use of the {\it Mathematica}$^{\cite{wolfram}}$ software in doing
the calculations and preparing the text of the present paper.
This work is supported in
part by the DOE grant DE-AC02-83ER40105.

\newpage

\unitlength=1.00mm
\linethickness{0.8pt}
\begin{picture}(130.00,95.00)
\put(85.00,95.00){\line(0,-1){99.00}}
\put(25.00,50.00){\line(1,0){105.00}}
\put(65.00,90.00){\line(0,-1){90.00}}
\put(85.00,15.00){\circle*{2.00}}
\put(45.00,15.00){\circle*{2.00}}
\put(125.00,15.00){\circle*{2.00}}
\put(90.00,94.00){\makebox(0,0)[cc]{Im {\it t}}}
\put(126.00,48.00){\makebox(0,0)[lt]{Re {\it t}}}
\put(66.00,94.00){\makebox(0,0)[cc]{Re $\tau$}}
\end{picture}

\vspace{0.5 in}
\begin{quote}
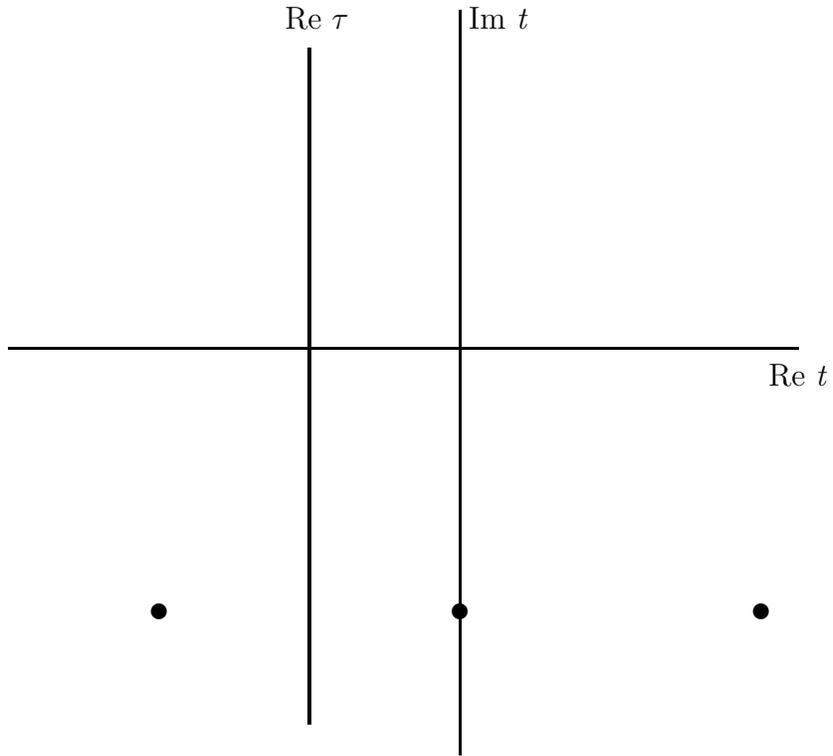
{\bf Fig.1} The structure of the classical field $\phi_0(t)$
(eq.(\ref{phi0})) in the complex $t$ plane. Heavy dots indicate the poles of
$\phi_0$. The vertical line going between the poles is the axis of real
$\tau$ on which the operator (\ref{oper}) is real.  \end{quote}

\end{document}